\documentclass{aip-cp}

\usepackage[numbers]{natbib}
\usepackage{rotating}
\usepackage{graphicx}

\begin{document}

\title{Large Size Telescope Report}

\author[aff1,aff2]{D. Mazin\corref{cor1}}
\author[aff3]{J.~Cortina}
\author[aff1,aff2]{M.~Teshima}
\author[]{and the CTA Consortium}
\eaddress{http://www.cta-observatory.org}

\affil[aff1]{Max-Planck-Institut f\"ur Physik, D-80805 M\"unchen, Germany.}
\affil[aff2]{Institute for Cosmic Ray Research, University of Tokyo, 277-8582 Chiba, Japan.}
\affil[aff3]{Institut de Fisica d'Altes Energies, The Barcelona Institute of Science and Technology, 08193 Bellaterra, Spain.}
\corresp[cor1]{Corresponding author: mazin@mpp.mpg.de}

\maketitle

\begin{abstract}
The Cherenkov Telescope Array (CTA) observatory
will be deployed over two sites in the two hemispheres. 
Both sites will be equipped with four Large Size Telescopes (LSTs), 
which are crucial to achieve the science goals of CTA in the 20-200 GeV energy range. 
Each LST is equipped with a primary tessellated mirror dish of 23\,m diameter, 
supported by a structure made mainly of carbon fibre reinforced plastic tubes and aluminum joints. 
This solution guarantees light weight (around 100 tons), 
essential for fast repositioning to any position in the sky in $<$20 seconds. 
The camera is composed of 1855 photomultiplier tubes and embeds the control, 
readout and trigger electronics. 
The detailed design is now complete and production of the first LST, which will serve 
as a prototype for the remaining seven, is ongoing. 
The installation of the first LST at the Roque de 
los Muchachos Observatory on the Canary island of La Palma (Spain) started in July 2016. 
In this paper we will outline the technical solutions adopted to fulfill the design requirements, 
present results of element prototyping and describe the installation and operation plans.
\end{abstract}

\section{INTRODUCTION}

\begin{figure}[h]
\centerline{\includegraphics[width=0.45\textwidth]{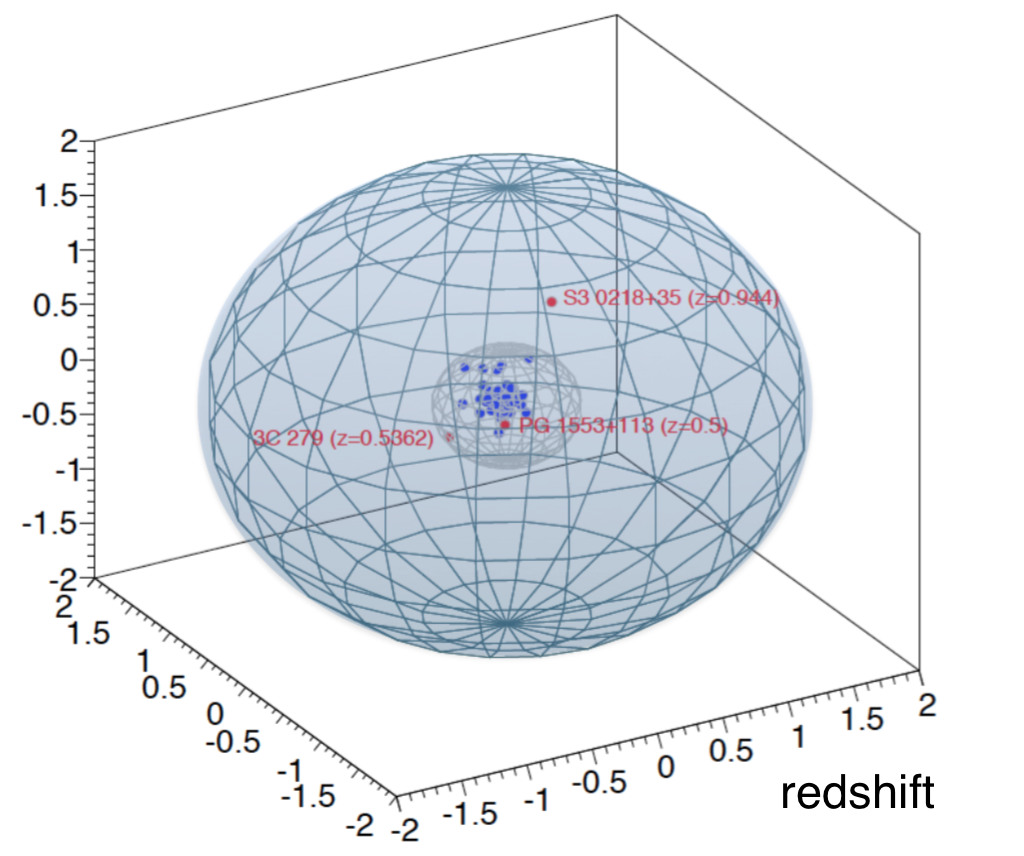}
\includegraphics[width=0.55\textwidth]{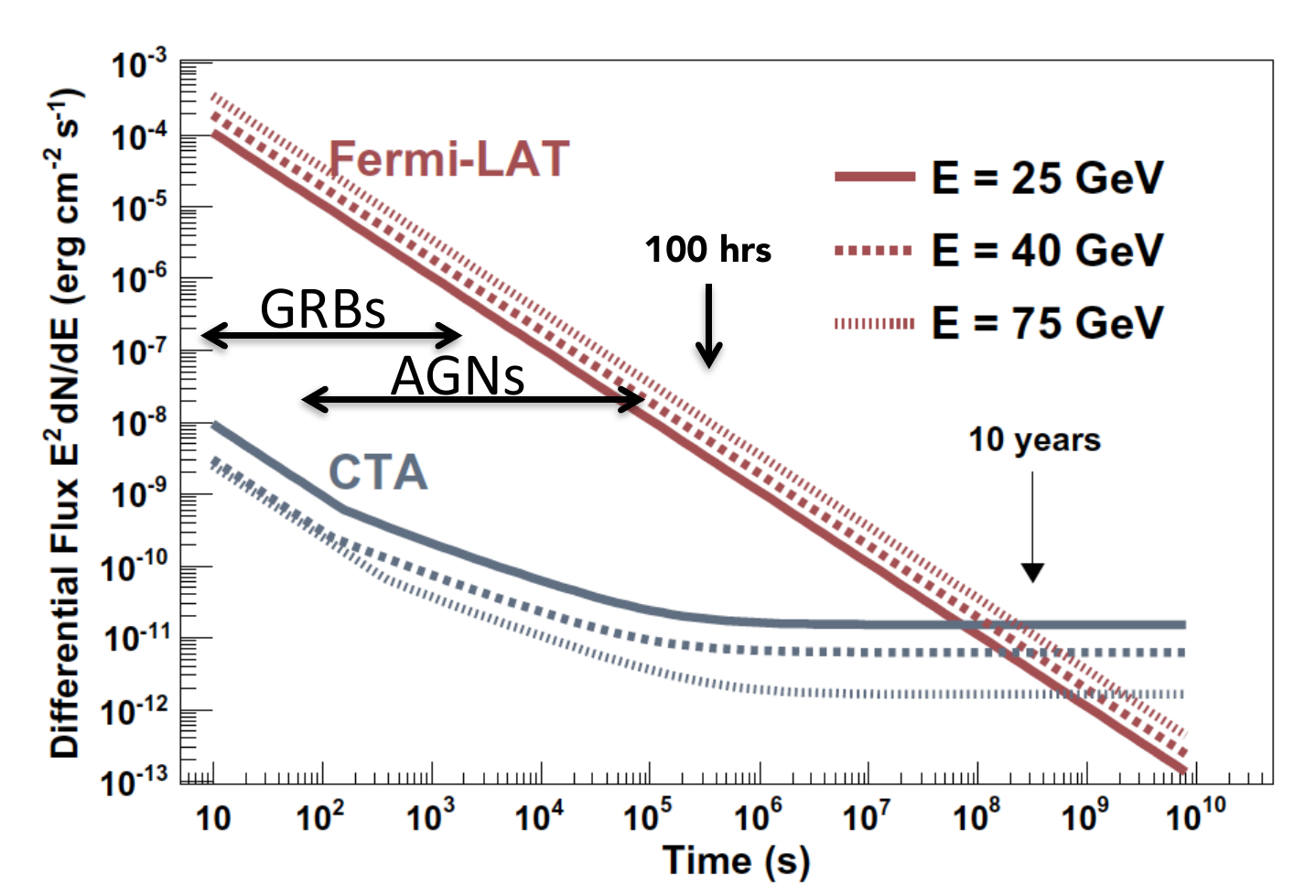}}
  \caption{The strengths of the LSTs within CTA: increase of the accessible volume in the universe (left plot,
the axes are in redshift units centered at the poistion of the Earth, in blue are the known TeV gamma-ray sources,
in red recently discovered ones),
and the time domain for transients and rapid variability phenomena (right).}
\label{fig:physics}
\end{figure}

The Large Size Telescopes (LSTs) of CTA are designed to cover the energy range from 20\,GeV to several TeV.
There will be 4 LSTs in CTA-South and CTA-North, each,
dominating the performance of the array between 20\,GeV and 200\,GeV.
The LST design targets the low energy threshold and the fast rotation speed
in order to a) provide good energy overlap with the currently existing space based instrument
of \emph{Fermi}-LAT; b) open up observable volume with Cherenkov telescopes from currently less than z=1
to more than z=2 (see Figure~\ref{fig:physics}, left plot); 
and c) study fast variability and transient phenomena such as flares of galactic and extragalactic sources
as well as observing GRBs
(see response to transient sources in Figure~\ref{fig:physics}, right plot, based on \cite{Funk2013a}).

In March 2015, the LST team and the IAC (Canary Islands) signed an agreement to build
the first LST (the prototype telescope that should become the first LST once successfully commissioned)
at the Observatorio del Roque de los Muchachos.
Since La Palma is the prime candidate to become CTA-North site\footnote{On 19 September 2016, the Council of the Cherenkov Telescope Array Observatory (CTAO) concluded negotiations with the Instituto de Astrofisica de Canarias (IAC) to host CTA’s northern hemisphere array at the Roque de los Muchachos Observatory in La Palma, Spain.}, it is a very convenient location,
meaning that the first LST will be built on its final site.

\section{OVERVIEW OF THE TECHNICAL DESIGN}

\begin{figure}[h!]
  \centerline{\includegraphics[width=0.75\textwidth]{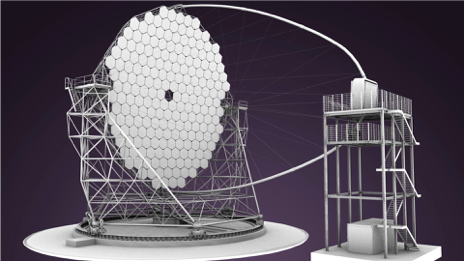}}
  \caption{The 3D model of the LST.}
\label{fig:LST}
\end{figure}

\begin{figure}[h!]
  \centerline{\includegraphics[width=0.75\textwidth]{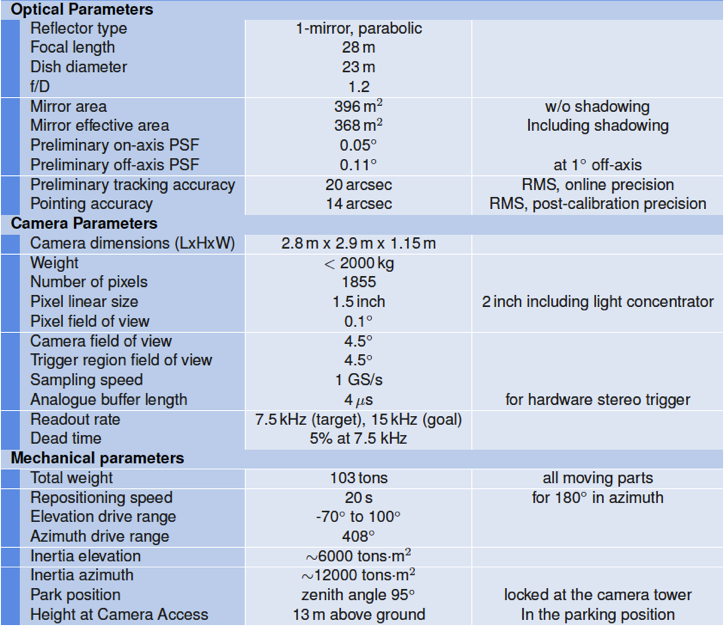}}
  \caption{The relevant technical parameters of the LST.}
\label{fig:params}
\end{figure}

The reader is referred to the Technical Design Report (TDR) of the LST for a
detailed description of the telescope and its component parts \citep{LST-TDR-2016a}. 
A short description of the main assemblies of LST can be found at \cite{Cortina2015a}.
Here we
only provide a general overview of the design as illustrated in Figure~\ref{fig:LST}. The
LST’s main parameters are summarized in tabular form in Figure~\ref{fig:params}.

%
\section{STATUS OF THE PRODUCTION}

The LST prototype is in the production phase. Most elements are produced at companies,
some assembly and elements are produced by participating institutes. In the following we
give a short status of the production.

\begin{figure}[h!]
\centerline{\includegraphics[width=0.205\textwidth]{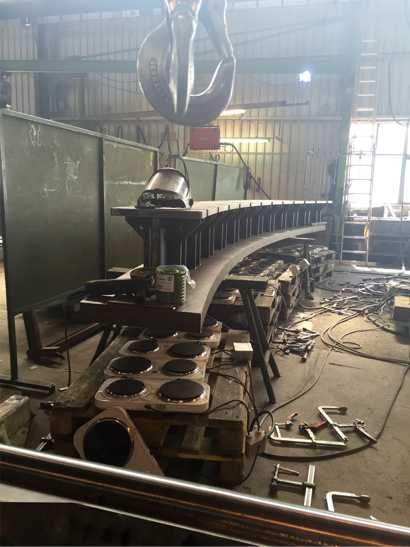}
\includegraphics[width=0.36\textwidth]{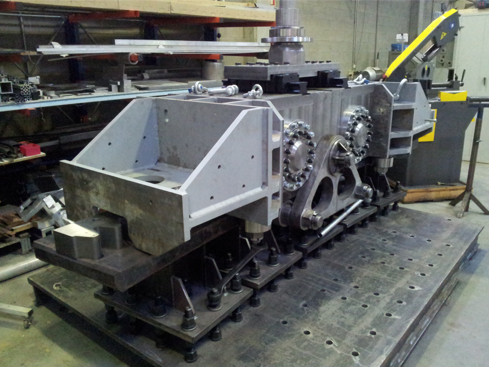}
\includegraphics[width=0.41\textwidth]{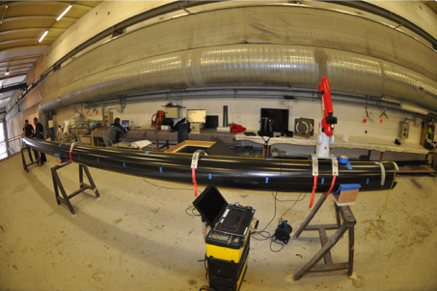}}
  \caption{Some produced parts of the LST structure. Left: Rail segment. Middle: one of the six Bogies.
Right: part of the Camera Support Structure.}
\label{fig:structure}
\end{figure}

\subsection{Telescope structure}

The Lower Structure of the LST is made of steel tubes; the Dish Structure is
from carbon fiber reinforced plastic (CFRP), steel and aluminium tubes.  The
Lower Structure and the Dish are being produced by the company
MERO-TSK\footnote{http://www.mero.de/index.php/en/}. The delivery is planned
for Autumn 2016 in La Palma.

The Lower Structure of the LST rests on
six Bogies equally spaced in a hexagonal arrangement, running on a circular
Rail. Two of those Bogies are located under the Elevation Bearings,
withstanding the higher percentage of the telescope’s weight. The Bogies run on
a circular flat Rail of 23.9 m diameter and 500 mm width, which is fixed to the
Foundation through the pedestals. One can see a segment of the Rail produced by
MBM Company in Germany\footnote{http://www.maschinenbau-muehldorf.de/en.html}
in Figure~\ref{fig:structure}, left picture. 
The six Bogies (see one of them in Figure~\ref{fig:structure}, middle picture) 
are produced in the LST member institutes, namely by
CIAMAT and IFAE (Spain), INFN Padova (Italy), University of Hamburg and MPI for physics (Germany).
IFAE in Barcelona coordinates the production and is responsible for the Bogie assembly on site.

The baseline design of the Camera Support
Structure is based on an almost parabolic arch geometry, reinforced along its
orthogonal projection by two symmetric sets of stabilizing fixed headstays.
Most of its elements make use of CFRP, which is well known to provide a very
high performance to mass ratio. The procurement is shared between LAPP in France and
INFN in Italy. See a segment (1 out of 6) of CSS in Figure~\ref{fig:structure}, right picture.

\subsection{Optical system}

\begin{figure}[h]
\centerline{\includegraphics[width=0.29\textwidth]{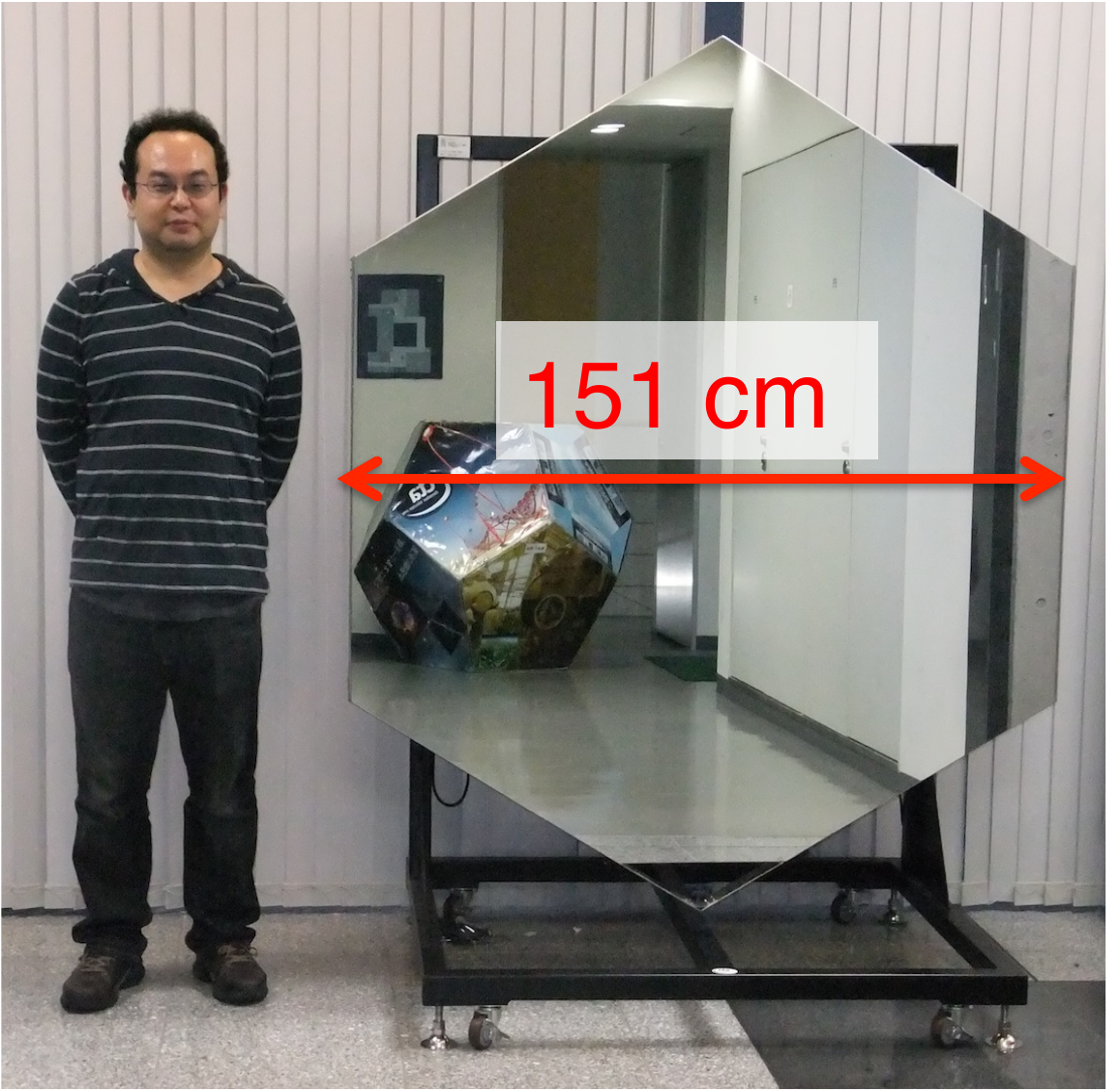}
\includegraphics[width=0.33\textwidth]{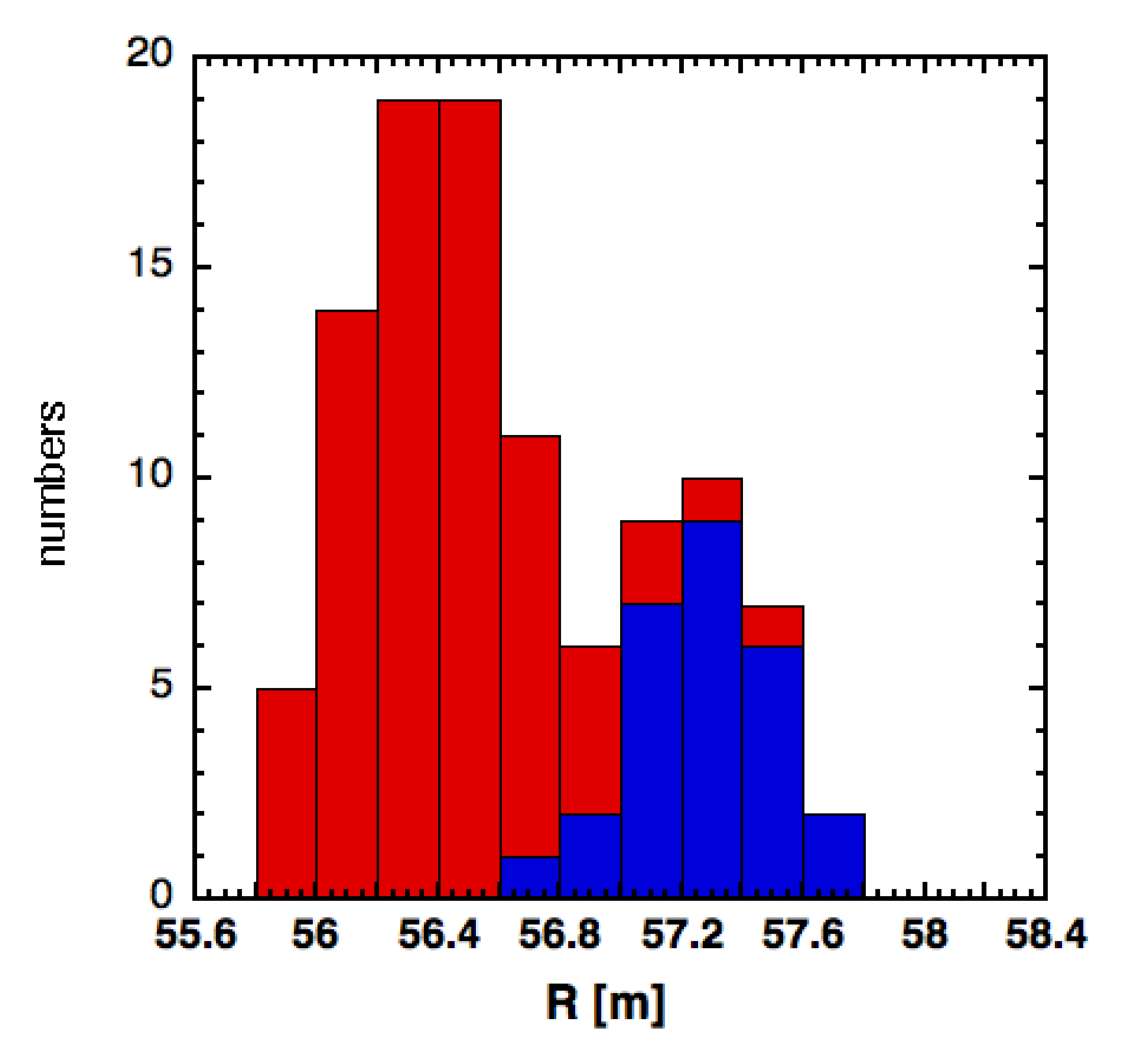}
\includegraphics[width=0.36\textwidth]{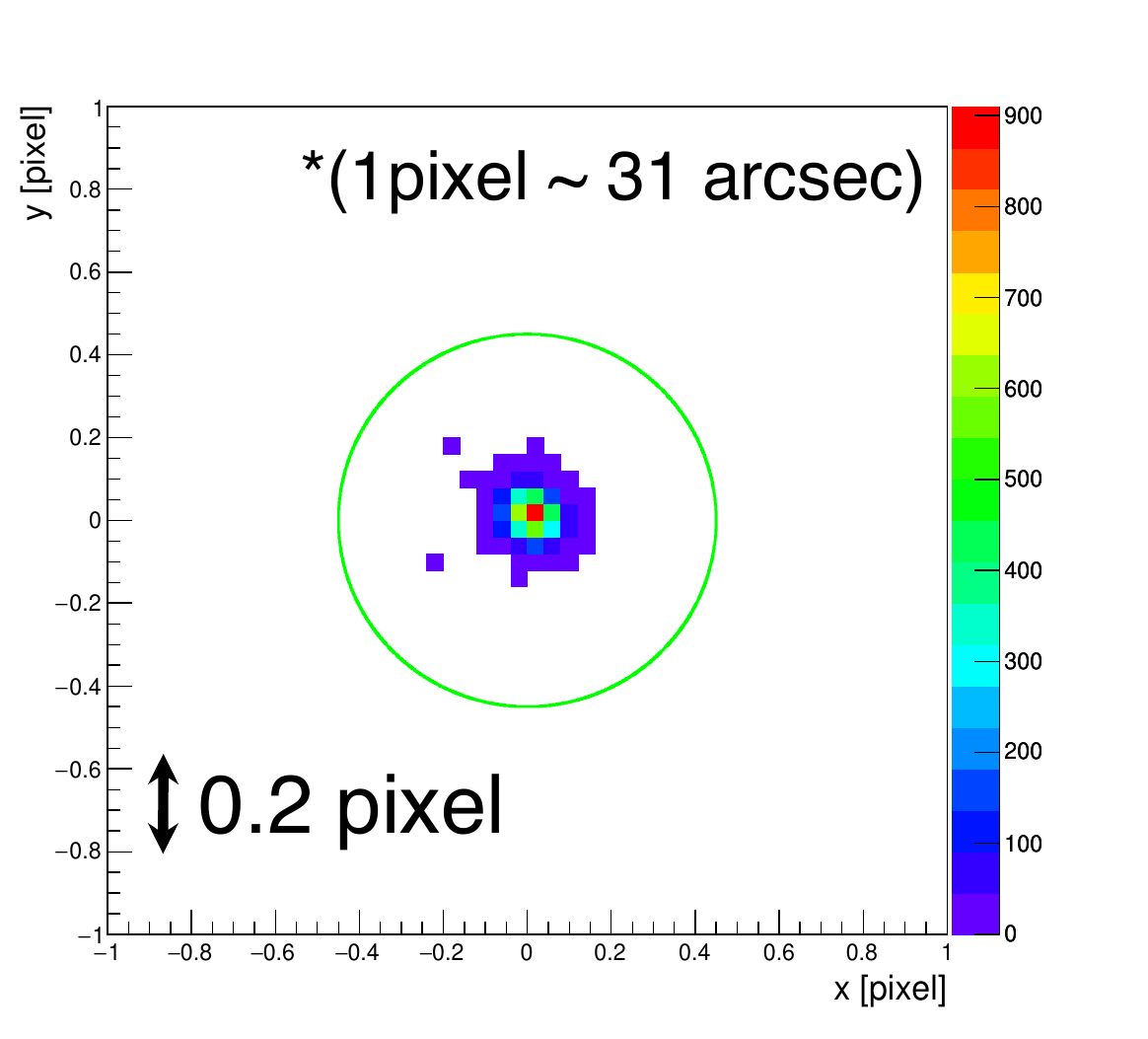}}
  \caption{Mirror and AMC system of LST. Left: Picture of a mirror. Middle: Measured radii from 60 mirrors delivered to ICRR from two different molds. Right: AMC performance to align a mirror 
from a random position to within the required green circle.}
\label{fig:mirrors}
\end{figure}

The Optical System of the LST is an active optics system that includes a large
parabolic reflector equipped with an Active Mirror Control system (find more
details in \cite{Hayashida2015a}) and a flat focal surface.

The optical reflector is composed of 198 hexagonal mirrors each with an area of
2m$^2$. The
Mirrors are manufactured using the cold slump technique with a sandwich
structure consisting of a soda-lime glass sheet, an aluminum honeycomb box and
another glass sheet. The mirror box is made of stainless-steel. 
The production of mirrors is ongoing in Japan by a company Sanko, 
60 are already produced, and 400 more will be produced in the next 12 months.
One can see one of them in Figure~\ref{fig:mirrors}, left plot. The middle panel shows
the measured mirrors radii made in two different moulds 
(marked with different colors, accordingly).
Due to the parabolic dish shape of LST we require three different moulds 
for the mirror production.

The mirrors are attached to the Dish of the LST structure using two actuators
and a fixed point. The actuators have accurate step motors ($5\,\mu\mbox{m}$ step size),
which are controlled by the Active Mirror Control (AMC) program to achieve the
required optical performance at any moment of time. Each mirror facet has a
small CMOS camera attached that observes a fixed-point (generated by a laser)
on the Camera plane, and the position of the fixed-point is used as a reference
to correct any misalignment of the mirrors. The actuators are produced by the University of Zurich.
In  Figure~\ref{fig:mirrors}, right plot, one can see the accuracy of actual alignment 
of a single mirrors using a reference spot and a CMOS camera.

\subsection{Camera}

\begin{figure}[h!]
\centerline{\includegraphics[width=0.46\textwidth]{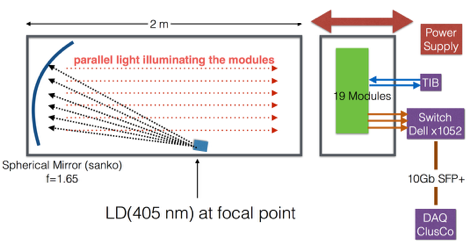}
\includegraphics[width=0.21\textwidth]{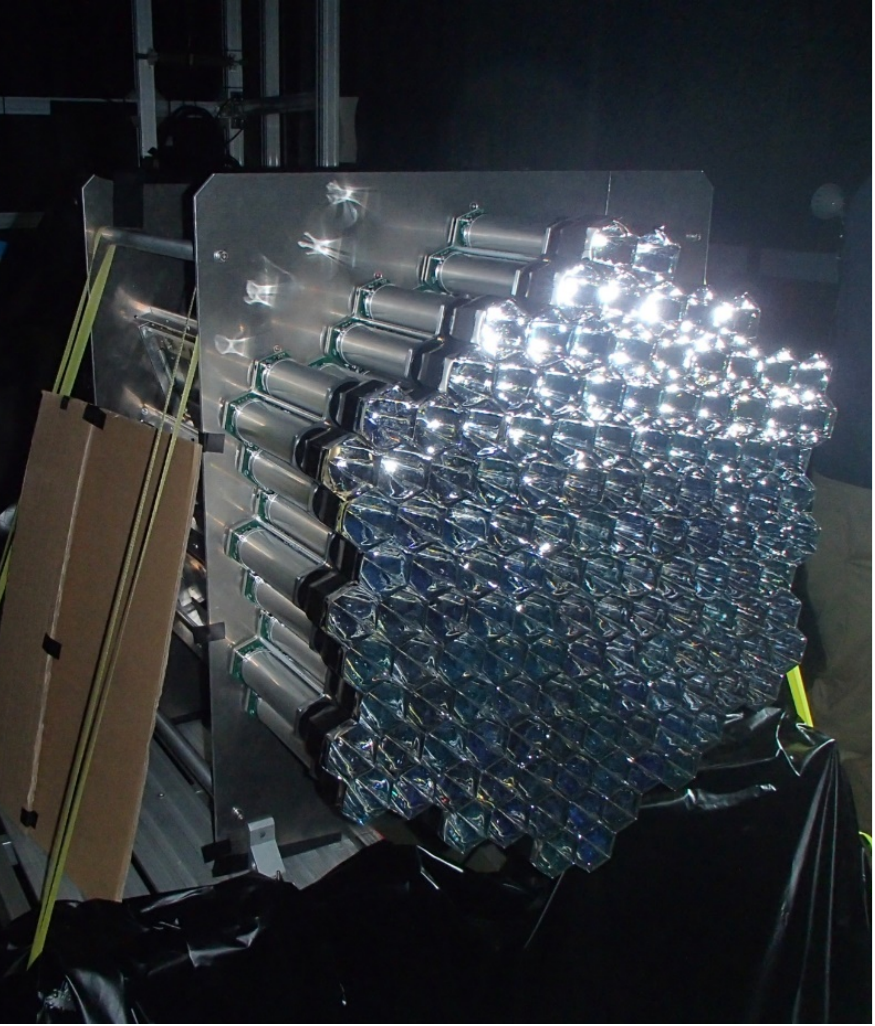}
\includegraphics[width=0.33\textwidth]{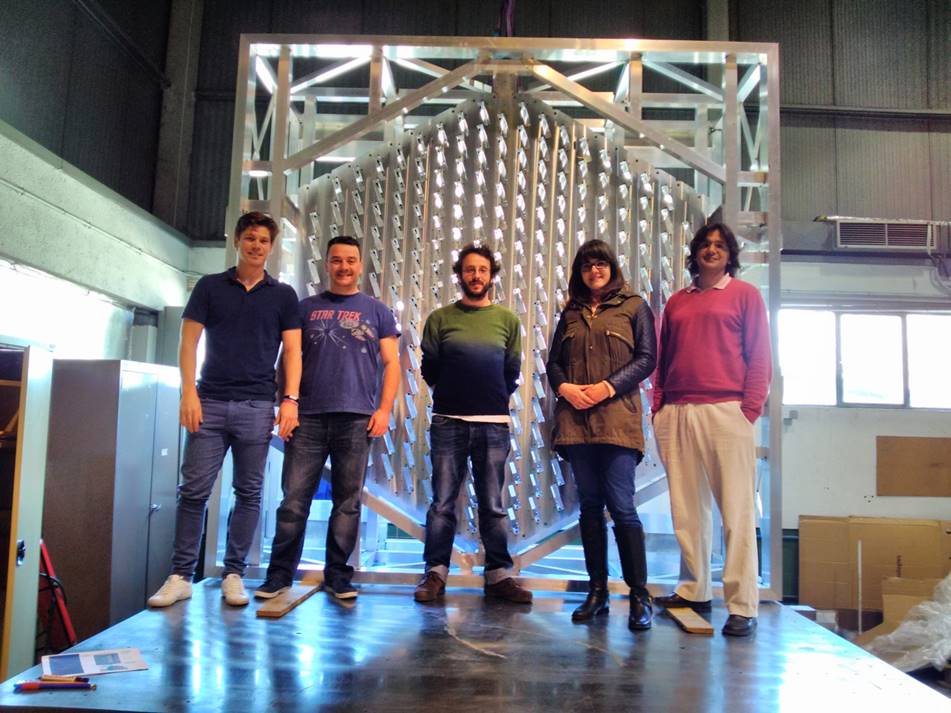}}
  \caption{Mini-Camera setup for 19 modules (left), Picture of the assembled Mini-Camera (middle) in Japan,
and the tubular structure of the entire LST camera (Spain).}
\label{fig:camera}
\end{figure}

The Camera of the LST shares many elements with the NectarCam \citep{Glicenstein2015a} 
proposed for use in the cameras on the Medium Size Telescopes of CTA. With a weight
of less than 2 tonnes the camera is comprised of 265 PMT modules 
that are easy to access and maintain. Each module has 7 channels, providing the
camera with a total of 1855 channels. Hamamatsu photomultiplier tubes (PMTs) with a
peak quantum efficiency of 42\% (R11920-100) are used as photosensors converting
the light to electrical signals that can be processed by dedicated electronics.
Each module incorporates a slow control board to steer and monitor the PMTs,
a trigger electronics (2 level) mounted on a mezzanine, a readout using DRS4 chips
as well as an analog backplane for trigger and clock propagation. 
A Mini-Camera with 19 modules is setup at ICRR, Japan to test and characterize 
all modules functionalities.
The schematic view of the Mini-Camera is shown on Figure~\ref{fig:camera},
a picture of the assembled Mini-Camera is shown in the central panel.
The Mini-Camera tests were successful and the production and quality control 
of the modules continues and should be finished by February 2017.

The camera mechanical structure is being produced at CIEMAT, Spain.
The tubular structure is ready (see Figure~\ref{fig:camera}, right panel),
and until end of 2016 all mechanical parts are expected to be finished and tested.
Then the camera mechanics and the modules will be shipped to IFAE in Barcelona 
for final integration and testing campaign before shipping the camera to La Palma.

\subsection{Auxiliary Systems}

The Auxiliary Systems refer to all instruments and devices on the LST excluding
the Camera. The main functions of these devices include driving the telescope,
correcting the telescope pointing or focus and calibrating the Camera. The
devices used for Structure Condition Monitoring as well as Lightning Protection
are also considered to be Auxiliary Systems.

The fast and precise movement of the LST is achieved by using electric
servomotors on both the elevation and azimuth axis. Four synchronized motors
are used for the azimuth axis and two for the elevation one. Synchronization of
the motors is managed directly by the Drive controller. 
The procurement of the drive motors and electronics is about to start.

The central facet position of the LST’s mirror is kept vacant to accommodate
several devices. These include: a) a Calibration Light Source box used to
calibrate the gain of the camera’s photosensors; two boxes are designed by Indian
and Italian groups and will be tested and compared until end of 2016; 
b) an Optical Axis Reference
Laser used by the AMC to define the optical axis of the telescope (already purchased by MPI, Munich); 
c) an Inclinometer used to measure the pointing elevation (already purchased by MPI, Munich); 
d) a starguider camera to
relate the pointing of the telescope with respect to the sky field (purchased and tested by the Swedish group); 
e) a Camera Displacement Monitor used to monitor any sagging of the Camera 
(designed and already partily purchased by a Croatian group); and f) Distance
Meters used to measure any displacement or rotation of the Camera along the
optical axis with respect to the Dish centre (already procured). This relatively sophisticated
pointing setup is necessary to meet the strict requirements defined for the
LST’s pointing (14 arcsec, post-calibration) and lightweight structure.

\section{STATUS ON SITE AND PLANS}

\begin{figure}[h]
\centerline{\includegraphics[width=0.45\textwidth]{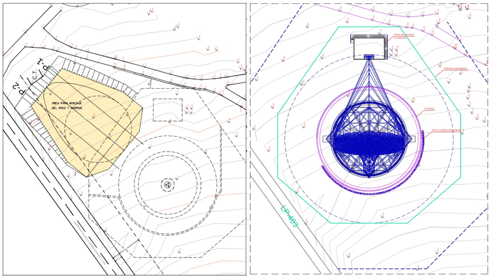}
\includegraphics[width=0.55\textwidth]{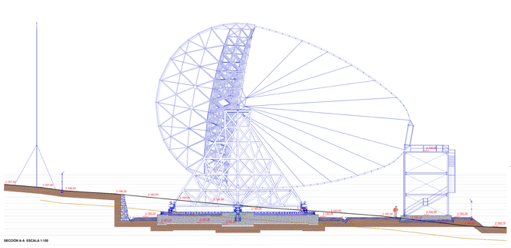}}
  \caption{Detailed design of the foundation and the mounting area for the LST prototype.}
\label{fig:site}
\end{figure}

The prototype LST is being built in La Palma.
Necessary permissions to start LST constructions in La Palma were obtained by May 2016 
and the foundation works started in July 2016. 
One can see detailed foundation and assembly area drawings in Figure~\ref{fig:site}.
The first LST will be positioned close to the existing MAGIC telescopes \citep{MAGICperformance20142}, 
which allows
one to cross-calibrate the new instrument with the existing ones.

The current planning foresees end of the foundation works by October 2016.
The Rail and the Bogies will be assembled then. After that, the Lower Structure will be assembled
on top of the Bogies. In parallel to the Lower Structure Assembly, the Dish Structure will be erected
next to that. Once ready (estimated time is 6 weeks), the Dish will be united with the lower structure.
Then the back part of the arch can be then mounted, safety access prepared and motors connected and cabled.
The telescope will be then moved to a parking position for easy access to the dish 
and mirrors as well as the AMC system mounted. This should take about 1.5 months.
After that the Telescope will be placed in vertical 
position\footnote{mirrors will be covered during CSS assembly}
and CSS (previously assembled on site) mounted on the dish. 
In the meanwhile the camera access tower must be ready. 
Then the telescope will be brought back to the parking position and the camera will be inserted
from the camera access tower. 
We plan to finish the installation by Summer 2017 and have the first light in October 2017.

The LST2-4 will be built in the CTA-North. The funding for them is basically secured, 
mainly through the Japanese and Spainish funding agencies 
with smaller contributions from France, Italy, Germany, India, Brazil, Sweden, and Croatia. 
The three telescopes can be built on La Palma in 2017-2019. The procurement for
CTA-South telescopes is expected to start in 2018. It can go in parallel with
the CTA-North telescopes.

\section{CONCLUSION}

The LST project as a part of CTA Consortium is progressing very well and construction of the first LST is ongoing.
The civil works for LST-1 started in La Palma in July 2016.
All telescope parts have been designed and are now in production and verification phases.
The construction of the first LST is expected to be finished by Summer 2017 and the construction
of the LST2-4 in La Palma by 2019. The 4 LSTs for CTA-South can be built and installed in 2018-2021,
according to the CTA plans.

\section{ACKNOWLEDGMENTS}
We gratefully acknowledge support from the agencies and organizations 
listed under Funding Agencies at this website: http://www.cta-observatory.org/.


\nocite{*}
\bibliographystyle{ieeetr}%
\bibliography{lstreport}%

\end{document}